\def\year{2021}\relax
\documentclass[letterpaper]{article} 
\usepackage{aaai21}  
\usepackage{times}  
\usepackage{helvet} 
\usepackage{courier}  
\usepackage[hyphens]{url}  
\usepackage{graphicx} 
\usepackage{natbib}  
\usepackage{caption} 
\usepackage[utf8]{inputenc}
\usepackage{graphicx}
\usepackage{amsmath}
\usepackage{amsthm}
\usepackage{booktabs}
\usepackage{algorithm}
\usepackage{algorithmic}
\usepackage{amssymb}
\usepackage{mathrsfs}
\usepackage{multirow}
\usepackage{array}
\usepackage{subfigure}
\DeclareUnicodeCharacter{00B1}{$\pm$}

\title{Adversarial Directed Graph Embedding}
\author {
    Shijie Zhu\textsuperscript{\rm 1,2},
    Jianxin Li\textsuperscript{\rm 1,2},
    Hao Peng\textsuperscript{\rm 1,2},
    Senzhang Wang\textsuperscript{\rm 3},
    Lifang He\textsuperscript{\rm 4} \\
}
\affiliations {
    \textsuperscript{\rm 1}Beijing Advanced Innovation Center for Big Data and Brain Computing, Beihang University, Beijing 100191, China \\
    \textsuperscript{\rm 2}State Key Laboratory of Software Development Environment, Beihang University, Beijing 100191, China\\
    \textsuperscript{\rm 3}College of Computer Science and Technology, Nanjing University of Aeronautics and Astronautics, Nanjing 211106, China  \\
    \textsuperscript{\rm 4}Department of Computer Science and Engineering, Lehigh University, Bethlehem, PA, USA \\
    \{zhusj,lijx,penghao\}@act.buaa.edu.cn, szwang@nuaa.edu.cn, lih319@lehigh.edu
}

\begin{document}

\maketitle

\begin{abstract}
Node representation learning for directed graphs is critically important to facilitate many graph mining tasks.
To capture the directed edges between nodes, existing methods mostly learn two embedding vectors for each node, source vector and target vector.
However, these methods learn the source and target vectors separately.
For the node with very low indegree or outdegree, the corresponding target vector or source vector cannot be effectively learned.
In this paper, we propose a novel Directed Graph embedding framework based on Generative Adversarial Network, called DGGAN.
The main idea is to use adversarial mechanisms to deploy a discriminator and two generators that jointly learn each node’s source and target vectors.
For a given node, the two generators are trained to generate its fake target and source neighbor nodes from the same underlying distribution, and the discriminator aims to distinguish whether a neighbor node is real or fake. 
The two generators are formulated into a unified framework and could mutually reinforce each other to learn more robust source and target vectors.
Extensive experiments show that DGGAN consistently and significantly outperforms existing state-of-the-art methods across multiple graph mining tasks on directed graphs.
\end{abstract}

\section{Introduction}

Graph embedding aims to learn a low-dimensional vector representation of each node in a graph, and has gained increasing research attention recently due to its wide and practical applications, such as link prediction~\cite{liben2007link}, graph reconstruction~\cite{tsitsulin2018verse}, node recommendation~\cite{ying2018graph}, and node classification~\cite{bhagat2011node}.

\begin{figure}[t]
\centering
\includegraphics[width=0.46\textwidth]{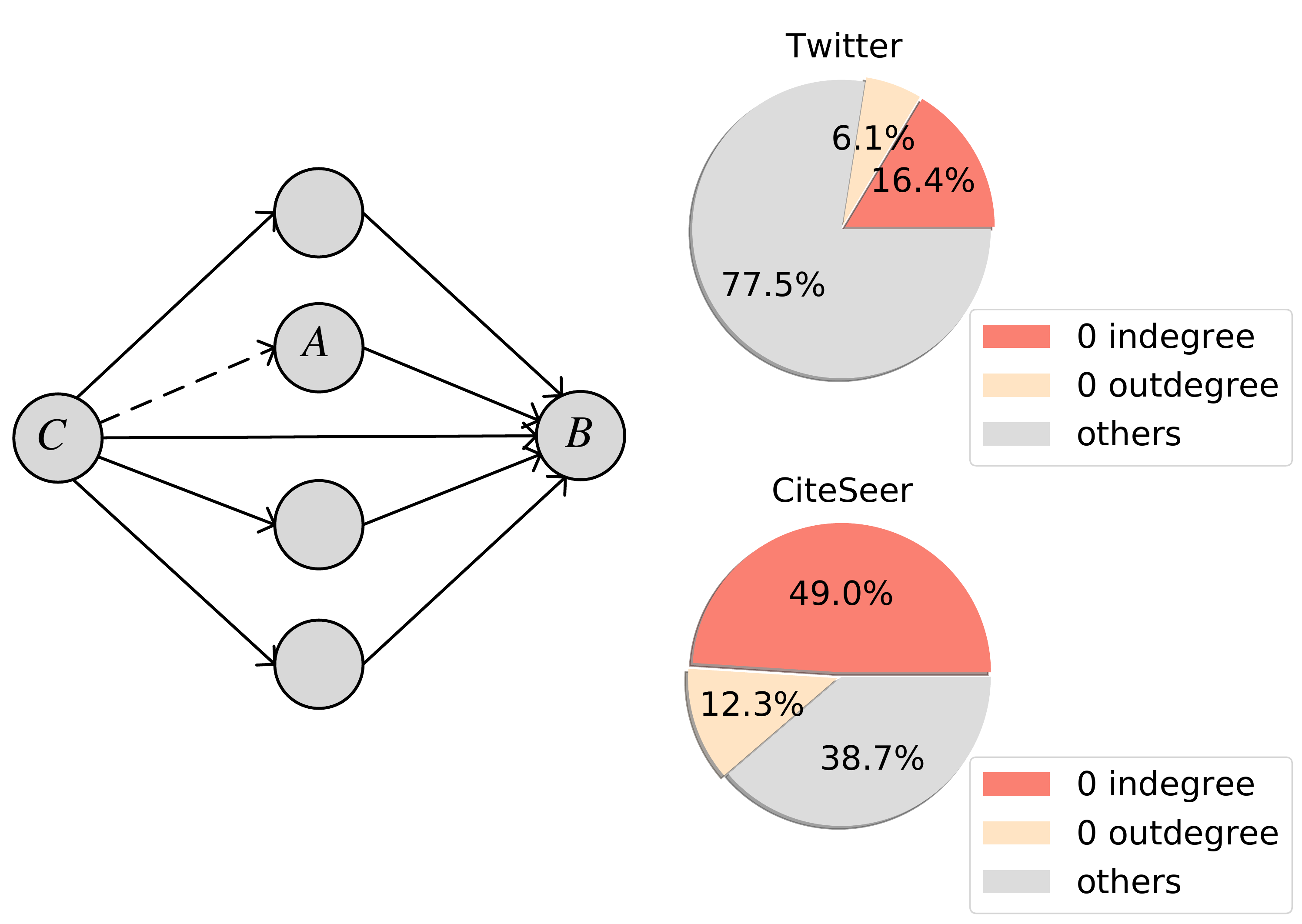}
\caption{The left figure is a toy example of directed graph. Considering the edge from $A$ to $B$, $B$ is the target neighbor of $A$ and $A$ is the source neighbor of $B$. The right two figures are statistics from the social network of Twitter \protect\cite{cha2010measuring} and the citation network of CiteSeer \protect\cite{bollacker1998citeseer}, respectively.}
\label{fig:DegreeRatio}
\end{figure}

Most existing methods, such as DeepWalk~\cite{perozzi2014deepwalk}, node2vec~\cite{grover2016node2vec}, LINE~\cite{tang2015line}, and GraphGAN~\cite{wang2018graphgan}, are designed to handle undirected graphs, but ignore the directions of the edges.
While the directionality often contains important asymmetric semantic information in directed graphs such as social networks, citation networks and webpage networks.
To preserve the directionality of the edges, some recent works try to use two node embedding spaces to represent the source role and the target role of the nodes, one corresponding to the outgoing direction and one for the incoming direction. 

We argue that there are two major limitations for existing directed graph embedding methods: 
(1) 
Methods like HOPE~\cite{ou2016asymmetric} rely on strict proximity measures like Katz~\cite{katz1953new} and low rank assumption of the graph.
Thus, they are difficult to be generalized to different types of graphs and tasks~\cite{khosla2018node}.
Moreover, HOPE is not scalable to large graphs as it requires the entire graph matrix as input and then adopts matrix factorization~\cite{tsitsulin2018verse}.
(2) Existing shallow methods focus on preserving the structure proximities but ignore the underlying distribution of the nodes.
Methods like APP~\cite{zhou2017scalable} adopt directed random walk sampling technique which follows the outgoing direction to sample node pairs.
These methods utilize negative sampling technique to randomly select existing nodes from the graph as negative samples.
However, for nodes with only outgoing edges or only incoming edges, the target or source vectors cannot be effectively trained. 
Figure~\ref{fig:DegreeRatio} presents a toy example.
Although both of the nodes $A$ and $C$ have no incoming edges, it is more likely to exist an edge from $C$ to $A$ than the other way round.
However, the proximities of the node pairs $(A, C)$ and $(C, A)$ predicted by APP are both zero, since the two node pairs are regarded as negative samples.
The several proximity measurements introduced by HOPE like Katz~\cite{katz1953new} all predict both proximities to be zero, as well.
As shown in Figure~\ref{fig:DegreeRatio}, the nodes with 0 indegree or 0 outdegree (e.g., $A$ and $B$) account for a large proportion of the graph.
The directed graph embedding methods mentioned above treat the source and target roles of each node separately, which causes these methods not robust.
However, a node's source role and target role are two types of properties of the node and are likely to be implicitly related. 
For instance, on social networks like Twitter, fans who follow a star may be followed by other fans with common interests.

In this paper, we propose DGGAN, a novel Directed Graph embedding framework based on Generative Adversarial Network (GAN)~\cite{goodfellow2014generative}.
Specifically, we train one discriminator and two generators which jointly generate the target and source neighborhoods for each node from the same underlying continuous distribution. 
Compared with existing methods, DGGAN generates fake nodes directly from a continuous distribution and is not sensitive to different graph structures.
Furthermore, the two generators are formulated into a unified framework and could naturally benefit from each other for better generations.
Under such framework, DGGAN could learn an effective target vector for the node $A$ in Figure~\ref{fig:DegreeRatio}, and will predict a high proximity for the node pair $(C, A)$.
The discriminator is trained to distinguish whether the generated neighborhood is real or fake. 
Competition between the generators and discriminator drives both of them to improve their capability until the generators are indistinguishable from the true connectivity distribution. 

The key contributions of this paper are as follows:
\begin{itemize}
    \item To the best of our knowledge, DGGAN is the first GAN-based method for directed graph embedding that could jointly learn the source vector and target vector for each node.
    \item The two generators deployed in DGGAN are also able to generate effective negative samples for nodes with low or zero out- or in- degree, which makes the model learn more robust node embeddings across various graphs. 
    \item Through extensive experiments on four real-world network datasets, we present that the proposed DGGAN method consistently and significantly outperforms various state-of-the-art methods on link prediction, node classification and graph reconstruction tasks.
\end{itemize}

\section{Related Work}

\subsection{Undirected Graph Embedding}
Graph embedding methods can be classified into three categories: matrix factorization-based models, random walk-based models and deep learning-based models. 
The matrix factorization-based models, such as GraRep~\cite{cao2015grarep} and M-NMF~\cite{wang2017community} first preprocess adjacency matrix which preserves the graph structure, and then decompose the prepocessed matrix to obtain graph embeddings.
It has been shown that many recent emergence random walk-based models such as DeepWalk~\cite{perozzi2014deepwalk}, LINE~\cite{tang2015line}, PTE~\cite{tang2015pte} and node2vec~\cite{grover2016node2vec} can be unified into the matrix factorization framework with closed forms~\cite{qiu2018network}. 
The deep learning-based models like SDNE~\cite{wang2016structural} and DNGR~\cite{cao2016deep} learn graph embeddings by deep autoencoder model.

\paragraph{Adversarial Graph Embedding}
Recently, Generative Adversarial Network (GAN)~\cite{goodfellow2014generative} has received increasing attention due to its impressing performance on the unsupervised task. 
GAN can be viewed as a minimax game between generator $G$ and discriminator $D$.
Formally, the objective function is defined as follows:
\begin{align}
\min\limits_{\theta^G }\max\limits_{\theta^D } \quad & \mathbb{E}_{x\sim p_{\rm data}(x)}\left [ \log D(x; \theta^D) \right ] \nonumber \\
+ & \mathbb{E}_{z\sim p_{z}(z)}\left [ \log (1 - D(G(z; \theta^G); \theta^D)) \right ] ,
\label{eq1}
\end{align}%
where $\theta^G$ and $\theta^D$ denote the parameters of $G$ and $D$, respectively.
The generator $G$ tries to generate close-to-real fake samples with the noise $z$ from a predefined distribution $p_{z}(z)$.
While the discriminator $D$ aims to distinguish the real ones from the distribution $p_{\rm data}(x)$ and the fake samples.
Several methods have been proposed to apply GAN for graph embedding to improve models robustness and generalization.
GraphGAN~\cite{wang2018graphgan} generates the sampling distribution to sample negative nodes. 
ANE~\cite{dai2018adversarial} imposes a prior distribution on graph embeddings through adversarial learning. 
NetRA~\cite{yu2018learning} and ARGA~\cite{pan2019learning} adopt adversarially regularized autoencoders to learn smoothly embeddings.
DWNS~\cite{dai2019adversarial} applies adversarial training by defining adversarial perturbations in embeddings space.

\subsection{Directed Graph Embedding}
The methods mentioned above mainly focus on undirected graphs and thus cannot capture the directions of edges.
There are some works for directed graph embedding, which commonly learn source embedding and target embedding for each node.
HOPE~\cite{ou2016asymmetric} derives the node-similarity matrix by approximating high-order proximity measures like Katz measure~\cite{katz1953new} and Rooted PageRank~\cite{song2009scalable}, and then decomposes the node-similarity matrix to obtain node embeddings. 
APP~\cite{zhou2017scalable} is a directed random walk-based method to implicitly preserve Rooted PageRank proximity.
NERD~\cite{khosla2018node} uses an alternating random walk strategy to sample node neighborhoods from a directed graph.
ATP~\cite{sun2019atp} incorporates graph hierarchy and reachability to construct the asymmetric matrix.
However, these methods are all shallow methods, failing to capture the highly non-linear property in graphs and learn robust node embeddings.

\begin{figure*}[t]
\centering
\includegraphics[width=0.92\textwidth]{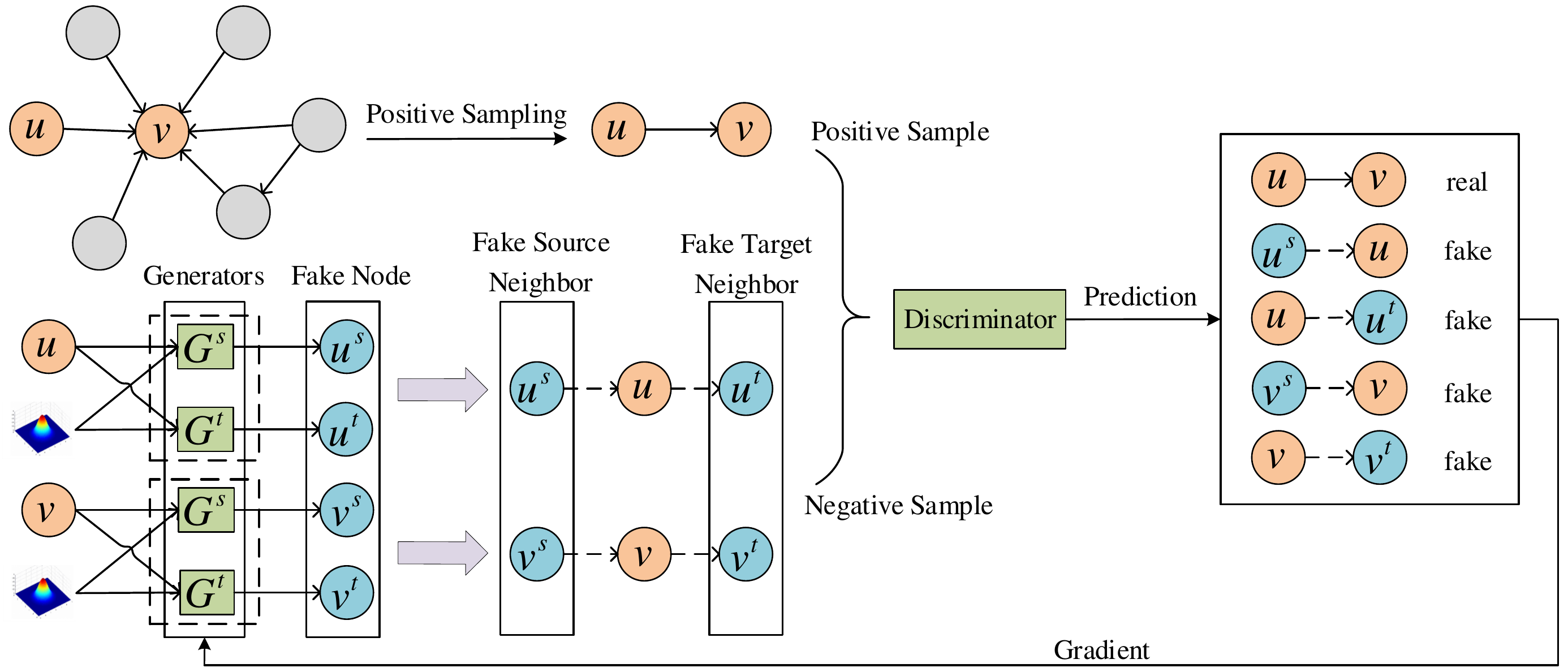}
\caption{The architecture of DGGAN. The node pair $(u, v)$ denotes real node pair. For node $u$, the two generators share an underlying distribution and jointly generate a fake source neighbor $u^s$ and a fake target neighbor $u^t$. Likewise, the fake source and target neighbors can be generated for node $v$. Those fake node pairs aim to fool the discriminator with highest probability, while the discriminator is trained to distinguish between the real node pair and fake node pair.}
\label{fig:Architecture}
\end{figure*}

\section{DGGAN Model}

In this section, we will first introduce the notations to be used. 
Then we will present an overview of DGGAN, followed by detailed descriptions of our generator and discriminator. 

\subsection{Notations}
We define a directed graph as $\mathcal{G} = \left \{ \mathcal{V}, \mathcal{E} \right \}$, where $\mathcal{V}$ is the node set, and $\mathcal{E}$ is the directed edge set.
For nodes $u, v \in \mathcal{V}$, $(u, v) \in \mathcal{E}$ represents a directed edge from $u$ to $v$.
To preserve the asymmetric proximity, each node $u$ needs to have two different roles, the source role and target role, represented by $d$ dimensional vector $\mathbf{s}_u \in \mathbb{R}^{d \times 1}$ and $\mathbf{t}_u \in \mathbb{R}^{d \times 1}$, respectively.

\subsection{Overall Framework of DGGAN}

The objective of DGGAN is to jointly learn the source and target vectors for each node on a directed graph.
Figure ~\ref{fig:Architecture} demonstrates the proposed framework of DGGAN, which mainly consists of two components: generator and discriminator.
Given a node (e.g., $u$), two generators are deployed to jointly generate its fake source neighborhood and target neighborhood from the same underlying continuous distribution.
And one discriminator is set to distinguish whether the source neighborhood and the target neighborhood of the given node are real or fake.
With the minimax game between the generator and discriminator, DGGAN is able to learn more robust source embeddings and target embeddings for nodes with low indegree or outdegree, even the nodes with zero indegree or outdegree (e.g., $u$ and $v$).
Next, we introduce the details of the generator and discriminator.

\subsection{Generator and Discriminator in DGGAN}

\subsubsection{Directed, Generalized and Robust Generator}
The goal of our generator $G$ is threefold: 
(1) It should generate close-to-real fake samples concerning specific direction.
Thus, given a node $u \in \mathcal{V}$, the generator $G$ aims to generate a fake source neighbor $u^s$ and a fake target neighbor $u^t$ where $u^s$ and $u^t$ should be as close as possible to the real nodes.
(2) It should be generalized to non-existent nodes.
In other words, the fake nodes $u^s$ and $u^t$ can be latent and not restricted to the original graph. 
(3) It should be able to generate efficient fake source and target neighborhoods for nodes with low or zero out- or in- degree.

To address the first aim, we design the generator $G$ consisting of two types of generators: one source neighborhood generator $G^s$ and one target neighborhood generator $G^t$.
For the second and third aims, we propose to introduce a latent variable $\mathbf{z} \sim p_{\mathbf{z}}(\mathbf{z})$ shared between $G^s$ and $G^t$ to generate samples.
Rather than directly generating samples from $p_{\mathbf{z}}(\mathbf{z})$, we integrate the multi-layer perception (MLP) into the generator for enhancing the expression of the fake samples, as deep neural networks have shown strong ability in capturing the highly non-linear property in a network~\cite{gao2019progan,DBLP:conf/kdd/HuFS19}.
Therefore, our generator $G$ is formulated as follows:
\begin{align}
G^{s}(u; \theta^{G^s}) = f^{s}(\mathbf{z}; \theta^{f^s}), \quad & G^{t}(u; \theta^{G^t}) = f^{t}(\mathbf{z}; \theta^{f^t}) , \nonumber \\
G(u; \theta^{G}) = \{ G^{s}(u;& \theta^{G^s}), G^{t}(u; \theta^{G^t})\} ,
\label{eq3}
\end{align}%
where $f^s$ and $f^t$ are implemented by MLP.
$\theta^{f^s}$ and $\theta^{f^t}$ denote the parameters of $f^s$ and $f^t$, respectively.
$\mathbf{z}$ serves as a bridge between $G^{s}$ and $G^{t}$.
With the help of $\mathbf{z}$, $G^s$ and $G^t$ are not independent, and can update each other indirectly to generate better fake source and target neighborhood.
Particularly, we derive $\mathbf{z}$ from the following Gaussian distribution:
\begin{align}
p_{\mathbf{z}}(\mathbf{z}) = \mathcal{N}(\mathbf{z}_{u}^{\rm T}, \sigma^{2}\mathbf{I}) ,
\label{eq2}
\end{align}%
where $\mathbf{z}_{u} \in \mathbb{R}^{d \times 1}$ is a learnable variable and stands for the latent representation of $u \in \mathcal{V}$.
The parameters of $G^s$ and $G^t$ are thus $\theta^{G^s} =  \{ \mathbf{z}_{u}^{\rm T}:u \in \mathcal{V}, \theta^{f^s} \}$ and $\theta^{G^t} = \{ \mathbf{z}_{u}^{\rm T}:u \in \mathcal{V}, \theta^{f^t} \}$, respectively.
Since $\theta^{G^s}$ and $\theta^{G^t}$ share parameter $\mathbf{z}_{u}^{\rm T}$, the parameters of the generator $G$ can be obtained as follows: 
\begin{align}
\theta^{G} = \{ \theta^{G^s}, \theta^{G^t} \} = \{ \mathbf{z}_{u}^{\rm T}:u \in \mathcal{V}, \theta^{f^s}, \theta^{f^t} \}.
\label{eq4}
\end{align}%

The generator $G$ aims to fool the discriminator $D$ by generating close-to-real fake samples.
To this end, the loss function of the generator is defined as follows:
%
\begin{equation}
\resizebox{.908\linewidth}{!}{$
    \displaystyle
\mathcal{L}^G = \mathbb{E}_{u \in \mathcal{V}} \log \left ( 1 - D(u^{s}, u) \right ) + \log \left ( 1 - D(u, u^{t}) \right ) ,
\label{eq5}
$}
\end{equation}
where $u^s$ and $u^t$ denote the fake source neighbor and fake target neighbor of $u$, respectively.
$D$ outputs the probability that the input node pair is real, and will be introduced in the next subsection. 
The source vector of $u^{s}$ and target vector of $u^{t}$ can be obtained by $G^s$ and $G^t$, i.e., $\mathbf{s}_{u^s} \sim G^{s}(u; \theta^{G^s}), \mathbf{t}_{u^t} \sim G^{t}(u; \theta^{G^t})$.
The parameters $\theta^{G}$ of the generator can be optimized by minimizing $\mathcal{L}^G$.

\subsubsection{Directed Discriminator}
The discriminator $D$ tries to distinguish the positive samples from the input graph $\mathcal{G}$ and the negative samples produced by the generator $G$.
Thus, $D$ could enforce $G$ to more accurately fit the real graph distribution $p_\mathcal{G}$.
Note that for a given node pair $(u, v)$, $D$ essentially outputs a probability that the sample $v$ is connected to $u$ in the outgoing direction.
For this purpose, we define the $D$ as the sigmoid function of the inner product of the input node pair $(u, v)$:
\begin{align}
D(u, v; \theta^{D}) = \frac{1}{1 + \exp(-\mathbf{s}_{u}^{\rm T} \cdot \mathbf{t}_v)} ,
\label{eq6}
\end{align}%
where $\theta^{D} = \{ \mathbf{s}_u, \mathbf{t}_u: u \in \mathcal{V} \}$ is the parameter for $D$, i.e., the union of source role embeddings and target role embeddings of all real nodes on the observed $\mathcal{G}$.
Specifically, the input node pair can be divided into the following two cases.

\paragraph{Positive Sample}
There indeed exists a directed edge from $u$ to $v$ on the $\mathcal{G}$, i.e., $(u, v) \in \mathcal{E}$, such as $(u, v)$ shown in Figure \ref{fig:Architecture}. 
Such node pair $(u, v)$ is considered positive and can be modeled by the following loss:
\begin{align}
\mathcal{L}_{\rm pos}^{D} = \mathbb{E}_{(u, v) \sim p_\mathcal{G}} - \log D(u, v) .
\label{eq7}
\end{align}%

\paragraph{Negative Sample}
For a given node $u \in \mathcal{V}$, $u^s$ and $u^t$ denote its fake source neighbor and fake target neighbor generated by $G^s$ and $G^t$, respectively, i.e., $\mathbf{s}_{u^s} \sim G^{s}(u; \theta^{G^s}), \mathbf{t}_{u^t} \sim G^{t}(u; \theta^{G^t})$, such as $(u^s, u)$ and $(u, u^t)$ shown in Figure \ref{fig:Architecture}.
Such node pairs $(u^s, u)$ and $(u, u^t)$ are considered negative and can be modeled by the following loss:
\begin{equation}
\resizebox{.908\linewidth}{!}{$
    \displaystyle
\mathcal{L}_{\rm neg}^{D} = \mathbb{E}_{u \in \mathcal{V}} - \log \left ( 1 - D(u^{s}, u) \right ) - \log \left ( 1 - D(u, u^{t}) \right ).
$} 
\label{eq8}
\end{equation}
Note that the fake node embedding $\mathbf{s}_{u^s}$ and $\mathbf{t}_{u^t}$ are not included in $\theta^{D}$ and the discriminator $D$ simply treats them as non-learnable input.

We integrate above two parts to train the discriminator:
\begin{align}
\mathcal{L}^{D} = \mathcal{L}_{\rm pos}^{D} + \mathcal{L}_{\rm neg}^{D} .
\label{eq9}
\end{align}%
The parameters $\theta^{D}$ of the discriminator can be optimized by minimizing $\mathcal{L}^{D}$.

\subsection{Training of DGGAN}

\paragraph{Training Algorithm}
In each training epoch, we alternate the training between the discriminator $D$ and generator $G$ with mini-batch gradient descent. 
Specifically, we first fix $\theta^G$ and the two generators jointly generate fake neighborhoods for each node pair on the graph to optimize $\theta^D$.
Then we fix $\theta^D$ and optimize $\theta^G$ to generate close-to-real fake neighborhoods for each node under the guidance of the discriminator $D$.
The discriminator and generator play against each other until DGGAN converges. 
The overall training algorithm for DGGAN is summarized in Algorithm \ref{alg:algorithm}.

\begin{algorithm}[tb]
\caption{DGGAN framework}
\label{alg:algorithm}
\textbf{Require}: directed graph $\mathcal{G}$, number of maximum training epochs $n^{epoch}$, numbers of generator and discriminator training iterations per epoch $n^G$, $n^D$, number of samples $n^s$ \\
\textbf{Ensure}: $\theta^G$, $\theta^D$
\begin{algorithmic}[1]
\STATE Initialize $\theta^G$ and $\theta^D$ for $G$ and $D$, respectively
\FOR{$epoch = 0; epoch < n^{epoch}$}
\FOR{$n = 0; n < n^D$}
\STATE Generate $n^s$ fake source neighbors $u^s, v^s$ and fake target neighbors $u^t, v^t$ for each node pair $(u, v) \in \mathcal{E}$
\STATE Update $\theta^D$ according to Eq.(\ref{eq9})
\ENDFOR
\FOR{$n = 0; n < n^G$}
\STATE Generate $n^s$ fake source neighbors $u^s$ and fake target neighbors $u^t$ for each node $u \in \mathcal{V}$
\STATE Update $\theta^G$ according to Eq.(\ref{eq5})
\ENDFOR
\ENDFOR
\end{algorithmic}
\end{algorithm}

\paragraph{Complexity Analysis}

The time complexity for the discriminator $D$ in each iteration is $O(n^s \cdot |\mathcal{E}| \cdot d^2)$, where $|\mathcal{E}|$ is the number of edges and $d$ is the embedding dimension of each node.
The time complexity for the generator $G$ in each iteration is $O(n^s \cdot |\mathcal{V}| \cdot d^2)$, where $|\mathcal{V}|$ is the number of nodes.
Therefore, the overall time complexity of DGGAN per epoch is $O \left(n^s \cdot (n^D \cdot |\mathcal{E}| + n^G \cdot |\mathcal{V}|) \cdot d^2 \right)$.
Since $n^s$, $n^G$, $n^D$ and $d$ are small constants, the time complexity is linear to the number of edges $|\mathcal{E}|$.
The space complexity of DGGAN is $O(d \cdot |\mathcal{V}| + |\mathcal{E}|)$.
We can see that, DGGAN is both time and space efficent and is scalable for large scale graphs.

\section{Experiments}

In this section, we conduct extensive experiments on several datasets to investigate the performance of DGGAN.

\begin{table*}[htbp]
\centering
\small
\begin{tabular}{ccccccccccccc}
\toprule
\multirow{2}{*}{method} & \multicolumn{3}{c}{Cora}                                     & \multicolumn{3}{c}{Twitter}                                  & \multicolumn{3}{c}{Epinions}                                 & \multicolumn{3}{c}{Google}                                   \\ \cline{2-13} 
                        & 0\%                & 50\%               & 100\%              & 0\%                & 50\%               & 100\%              & 0\%                & 50\%               & 100\%              & 0\%                & 50\%               & 100\%              \\ 
\midrule
DeepWalk                & 84.9 & 68.1          & 52.9          & 50.4          & 50.3          & 50.3          & 76.6          & 67.9          & 66.6          & 83.6          & 72.1          & 66.5          \\
LINE-1                  & 84.7          & 68.0          & 52.5          & 53.1          & 51.5          & 50.0          & 78.8          & 69.8          & 68.5          & 89.7          & 72.7          & 65.1          \\
node2vec                & \textbf{85.3}          & 65.5          & 52.1          & 50.6          & 50.5          & 50.3          & 89.7          & 74.6          & 72.6          & 84.3          & 70.5          & 64.3          \\ 
\midrule
GraphGAN                & 51.6          & 51.3          & 51.2          & -                  & -                  & -                  & -                  & -                  & -                  & 71.3          & 61.1          & 56.2          \\
ANE                     & 72.8          & 61.4          & 51.5          & 49.7          & 49.8          & 50.0          & 85.5          & 69.2          & 66.9          & 76.1          & 63.7          & 57.8          \\ 
\midrule
LINE-2                  & 69.3          & 72.1          & 74.3          & 95.6          & 95.7          & 95.8          & 58.1          & 67.1          & 68.4          & 77.4          & 85.2          & \underline{89.0}    \\
HOPE                    & 77.6          & 74.2          & 71.5          & 98.0          & 97.9          & 97.8          & 79.6          & 71.7          & 70.5          & 87.5          & 85.6          & 84.6          \\
APP                     & 76.6          & 76.4          & 76.2          & 71.6          & 70.1          & 68.7          & 70.5          & 71.3          & 71.4          & \underline{92.1}    & 86.4          & 81.0          \\ 
\midrule
DGGAN*                  & 83.0          & \underline{83.3}    & \underline{83.5}    & \underline{99.4}    & \underline{99.3}    & \underline{99.2}    & \underline{92.7}    & \underline{80.0}    & \underline{78.2}    & 91.6          & \underline{89.2}    & 87.7          \\
DGGAN                   & \underline{85.1}    & \textbf{86.7} & \textbf{88.3} & \textbf{99.7} & \textbf{99.7} & \textbf{99.7} & \textbf{96.1} & \textbf{86.4} & \textbf{85.1} & \textbf{92.3} & \textbf{93.4} & \textbf{94.4} \\ 
\bottomrule
\end{tabular}
\caption{Area Under Curve (AUC) scores of link prediction on directed graphs with different fractions of positive edges except bi-directional edges been reversed to create negative edges in the test set (scores are with \%). The best scores are shown in bold and the second-best scores are underlined.}
\label{link_prediction}
\end{table*}

\begin{table}[htbp]
\centering
\begin{tabular}{@{}ccccc@{}}
\toprule
dataset  & \#nodes & \#edges & Avg. degree & \#labels \\ \midrule
Cora     & 23,166  & 91,500  & 7.90        & 10       \\
CoCit    & 44,034  & 195,361 & 8.86        & 15       \\
Twitter  & 465,017 & 834,797 & 3.59        & -        \\
Epinions & 75,879  & 508,837 & 13.41       & -        \\
Google   & 15,763  & 171,206 & 21.72       & -        \\ \bottomrule
\end{tabular}
\caption{Statistics of datasets.}
\label{dataset_description}
\end{table}

\subsection{Dataset Description}
We use four different types of directed graphs, including citation network, social network, trust network and hyperlink network to evaluate the performance of the model.
The details of the data are described as follows:
\textbf{Cora}~\cite{Cora} and \textbf{CoCit}~\cite{tsitsulin2018verse} are citation networks of academic papers. Nodes represent papers and directed edges represent citation relationships between papers. Labels represent conferences in which papers were published.
\textbf{Twitter}~\cite{Twitter} is a social network. Nodes represent users and directed edges represent following relationships between users.
\textbf{Epinions}~\cite{Epinions} is a trust network from the online social network Epinions. Nodes represent users and directed edges represent trust between users.
\textbf{Google}~\cite{Google} is a hyperlink network from pages within Google's sites. Nodes represent pages and directed edges represent hyperlink between pages.
The statistics of these networks are summarized into Table \ref{dataset_description}.

\subsection{Experiment Settings}

\begin{figure*}[ht]
\centering
\includegraphics[width=0.98\textwidth]{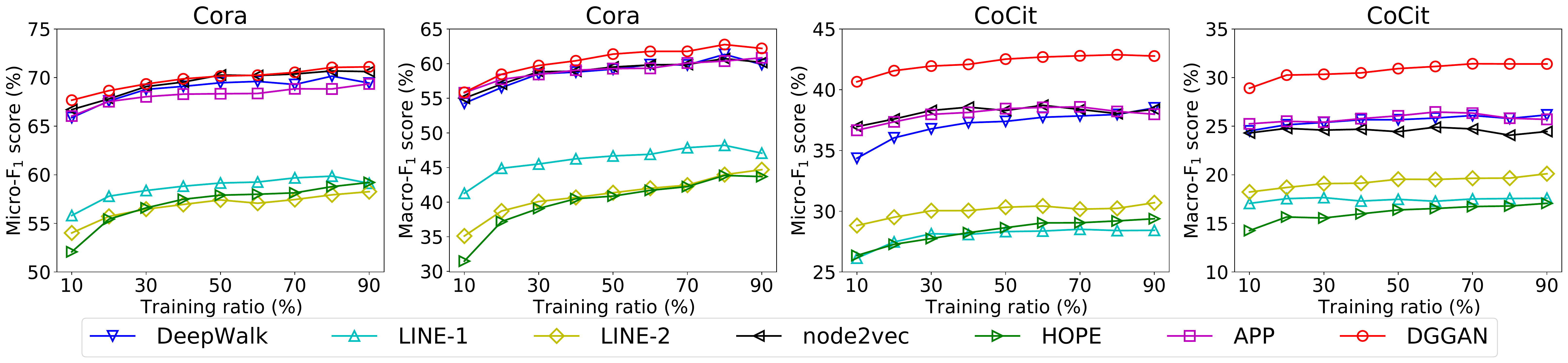}
\caption{Performance comparison of node classification task on Cora and CoCit datasets. The $x$ axis denotes the training ratio of labeled nodes, and the $y$ axis denotes Micro-F$_1$ or Macro-F$_1$ score.}
\label{figure3}
\end{figure*}

To verify the performance of DGGAN\footnote{https://github.com/RingBDStack/DGGAN}, we compare it with several state-of-the-art methods.

\begin{itemize}
    \item \textbf{Traditional undirected graph embedding methods}:
DeepWalk~\cite{perozzi2014deepwalk} uses local information obtained from truncated random walks to learn node embeddings.
LINE~\cite{tang2015line} learns large-scale information network embedding using first-order and second-order proximities.
node2vec~\cite{grover2016node2vec} is a variant of DeepWalk and utilizes a biased random walk algorithm to more efficiently explore the neighborhood architecture.

    \item \textbf{GAN-based undirected graph embedding methods}:
GraphGAN~\cite{wang2018graphgan} generates the sampling distribution to sample negative nodes from the graph.
ANE~\cite{dai2018adversarial} proposes to train a discriminator to push the embedding distribution to match the fixed prior.

    \item \textbf{Directed graph embedding methods}:
HOPE~\cite{ou2016asymmetric} preserves the asymmetric role information of the nodes by approximating high-order proximity measures.
APP~\cite{zhou2017scalable} proposes a random walk based method to encode Rooted PageRank proximity. 

    \item DGGAN* is a simplified version of DGGAN which uses only one generator $G^t$ to generate target neighborhoods of each node. 
We omit another simplified version which uses only one generator $G^s$ as we do not observe a significant performance difference compared with DGGAN*. 
\end{itemize}

For DeepWalk, node2vec and APP, the number of walks, the walk length and the window size are set to 10, 80 and 10, respectively, for fair comparision.
node2vec is optimized with grid search over its return and in-out parameters $(p, q) \in \{0.25, 0.5, 1, 2, 4\}$ on each dataset and task.
For LINE, we utilize both the first-order and the second-order proximities.
For the second-order proximities, node embeddings are considered as source embeddings, and context embeddings are used as target embeddings.
In addition, the number of negative samples is empirically set to 5. 
For GraphGAN, ANE and HOPE, we follow the parameters settings in the original papers.
Note that we do not report the results of GraphGAN on Twitter and Epinions datasets, since it cannot run on these two large datasets.
For DGGAN* and DGGAN, we choose parameters by cross validation and we fix the numbers of generator and discriminator training iterations per epoch $n^G=5, n^D=15$ across all datasets and tasks.
Throughout our experiments, the dimension of node embeddings is set to 128.


\subsection{Link Prediction}
In link prediction task, we predict missing edges given a network with a fraction of removed edges. 
A fraction of edges is removed randomly to serve as test split while the remaining network are utilized for training.
When removing edges randomly, we make sure that no node is isolated to avoid meaningless embedding vectors.
Specifically, we remove 50\% of edges for Cora, Epinions and Google datasets, and 40\% of edges for Twitter dataset.
Note that the test split is balanced with negative edges sampled from random node pairs that have no edges between them.
Since we are interested in both the existence of the edge between two nodes and the direction of the edge, we reverse a fraction of node pairs in the positive samples to replace the original negative samples if the edges are not bi-directional.
A value in $(0; 1]$ determines what fraction of positive edges from the test split are inverted at most to create negative examples.
And a value of 0 corresponds to the classical undirected graph setting where all the negative edges are sampled from random node pairs.

We summarize the Area Under Curve (AUC) scores for all methods in Table \ref{link_prediction}.
The reported results are the average performance of 10 times experiments.
Note that some methods like DeepWalk which mainly focus on undirected graphs, also achieve good performance on Cora dataset with random negative edges in test set.
But their performance decreases rapidly with the increase of reversed positive edges as they cannot model the asymmetric proximity, and their AUC scores are near 0.5 as expected. 
HOPE shows good performance on Twitter dataset but does not perform well on other datasets like Cora and Epinions.
It suggests that HOPE is difficult to be generalized to different types of graphs as mentioned above.
Note that on Epinions dataset, up to 31.5\% nodes have no incoming edges and 20.5\% nodes have no outgoing edges.
The directed graph embedding methods like APP show poor performance on Epinions dataset.
The reason is that these methods treat the source role and target role of one node separately, which renders them not robust.
We can see that DGGAN* shows much better performance than HOPE and APP across datasets.
This is because the negative samples of DGGAN* are generated directly from a continuous distribution and thus DGGAN* is not sensitive to different graph structures. 
Moreover, DGGAN outperforms DGGAN* as DGGAN utilizes two generators which mutually update each other to learn more robust source and target vectors.
Compared with baselines, the performance of DGGAN does not change much with different fractions of reversed positive test edges.
Overall, DGGAN shows more robustness across datasets and outperforms all methods in link prediction task.

\subsection{Node Classification}
To further verify whether a network embedding method can discover and preserve the proximity, we conduct multi-class classification on Cora and CoCit datasets.
Specifically, we randomly sample a fraction of the labeled nodes as training data and the task is to predict the labels for the remaining nodes. 
We train a standard one-vs-rest L2-regularized logistic regression classifier on the training data and evaluate its performance on the test data. 
We report Micro-F$_1$ and Macro-F$_1$ scores as evaluation metrics.
Note that for the methods using two embedding matrices, we set the dimension of node embeddings $d=64$ and concatenate the two 64-dimensional embedding vectors into a 128-dimensional vector to represent each node. 

Figure \ref{figure3} summarizes the experimental results when varying the training ratio of the labeled nodes.
Each result is averaged by 10 runs.
The results exhibit similar trends as follows.
First, DGGAN consistently outperforms all of the other methods across all training ratios on both datasets.
It demonstrates that DGGAN can effectively capture the neighborhood information in a robust manner through the adversarial learning framework.
Second, the undirected graph embedding methods DeepWalk and node2vec show good performance, and perform as well as the directed method APP.
This suggests that the directionality might have limited impact on performance for node classification task on these two datasets.
Third, we notice that although HOPE have good link prediction results, yet it performs poorly on this node classification task.
The reason might be that HOPE is linked to a particular proximity measure, which makes it hard to generalize to different tasks.

\subsection{Graph Reconstruction}

\begin{figure}[tbp]
\centering
\includegraphics[width=0.47\textwidth]{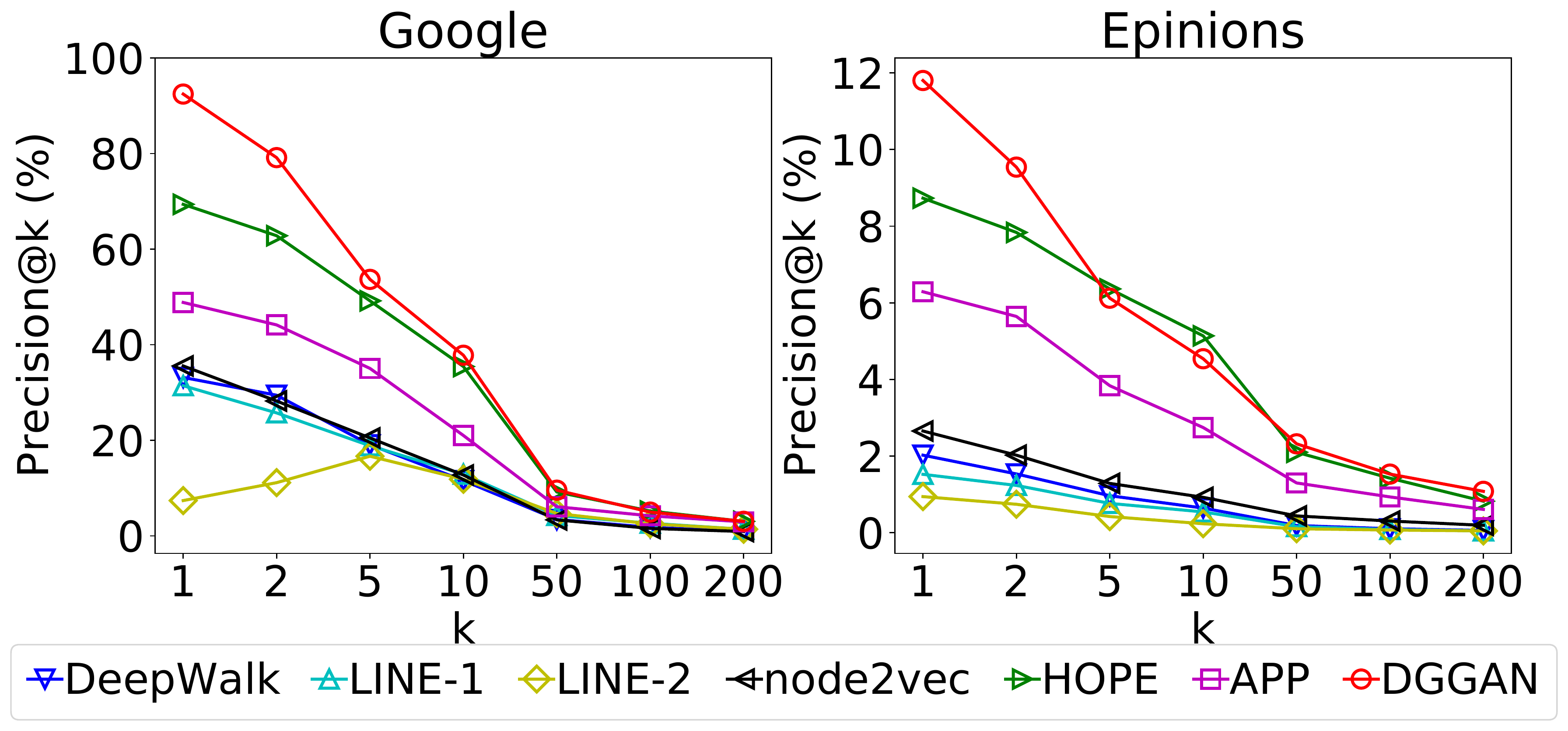}
\caption{Precision@k of graph reconstruction task on Google and Epinions datasets.}
\label{figure4}
\end{figure}

As the effective representations of a graph, node embeddings maintain the edge information and are expected to well reconstruct the original graph. 
We reconstruct the graph edges based on the reconstructed proximity between nodes.
Since each two adjacent nodes should be close in the embedding space, we use inner product between node vectors to reconstruct the proximity matrix.
For a given $k$, we obtain the $k$-nearest target neighbors ranked by reconstructed proximity for each method.
We perform the graph reconstruction task on Google and Epinions datasets.
In order to create the test set, we randomly sample 10\% of the nodes on each graph. 

We plot the average precision corresponding to different values of $k$ in Figure \ref{figure4}.
The results show that for both datasets, DGGAN outperforms baselines including HOPE and APP, especially when $k$ is small.
For Google dataset, DGGAN shows an improvement of around 33\% for $k=1$ over the second best performing method, HOPE. 
This shows the benefit of jointly learning the source and target vectors for each node. 
Some of the methods that focus on undirected graphs like node2vec exhibited good performance in link prediction.
However, these methods show poor performance in graph reconstruction.
This is because this task is harder than link prediction as the model needs to distinguish between small number of positive edges with a large number of negative edges.
Besides, we note that all the precision curves converge to points with small values when $k$ becomes large since most of the real target neighborhoods have been correctly predicted by these methods.

\subsection{Model Analysis}

In this subsection, we analyze the performance of different models under different levels of sparsity of networks and the converging performance of DGGAN. 
We choose Google dataset as it is much denser than the others. 
We first investigate how the sparsity of the networks affects the three directed graph embedding methods HOPE, APP and DGGAN. 
The setting of training procedure in this experiment is the same as link prediction and 50\% positive edges of test set are reversed to form negative edges.
We randomly select different ratios of edges from the original network to construct networks with different levels of sparsity.

\begin{figure}[tbp]
\centering
\subfigure[Sparsity]{\includegraphics[width=0.225\textwidth]{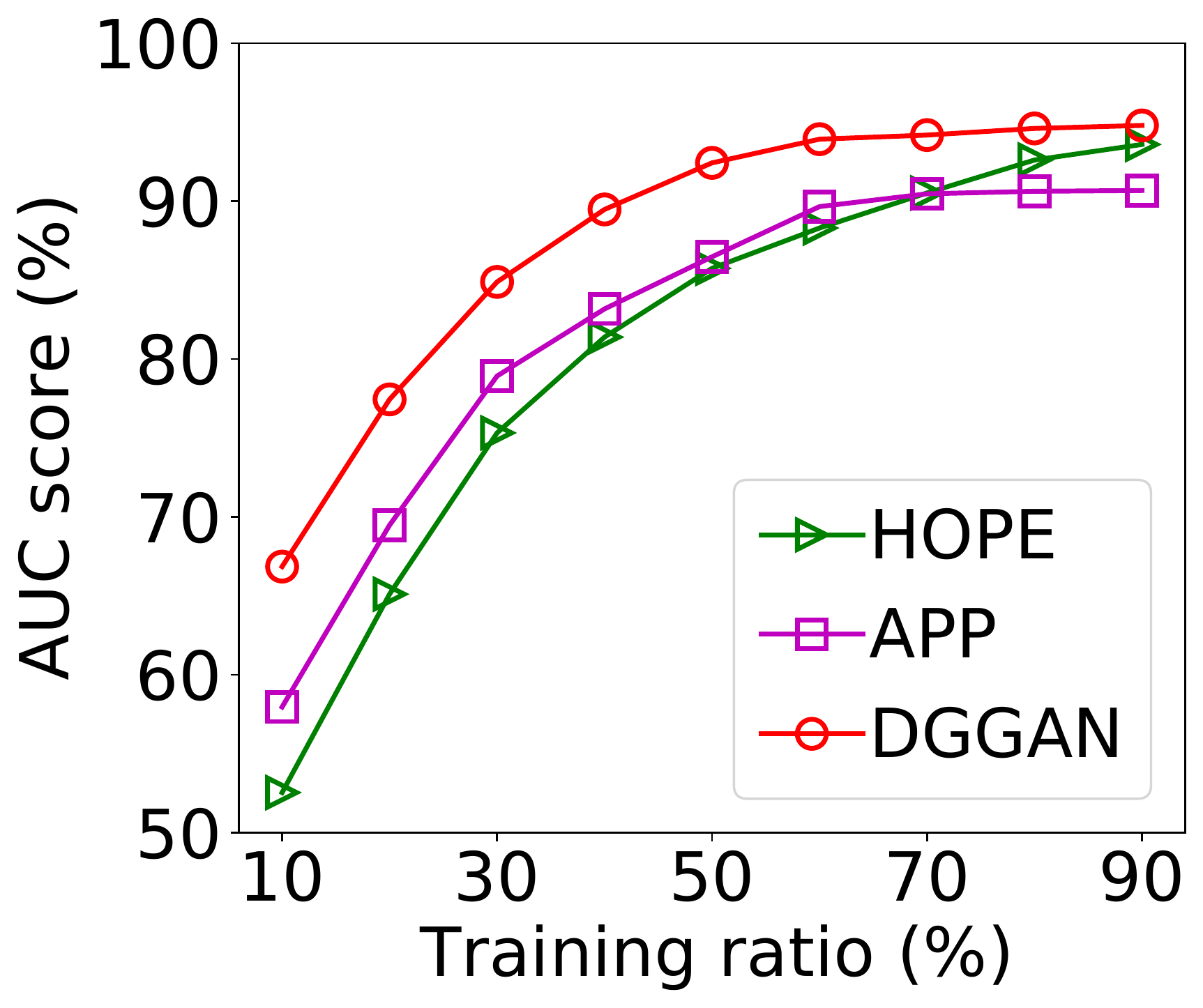} \label{figure5-a}}
\subfigure[Learning curves]{\includegraphics[width=0.225\textwidth]{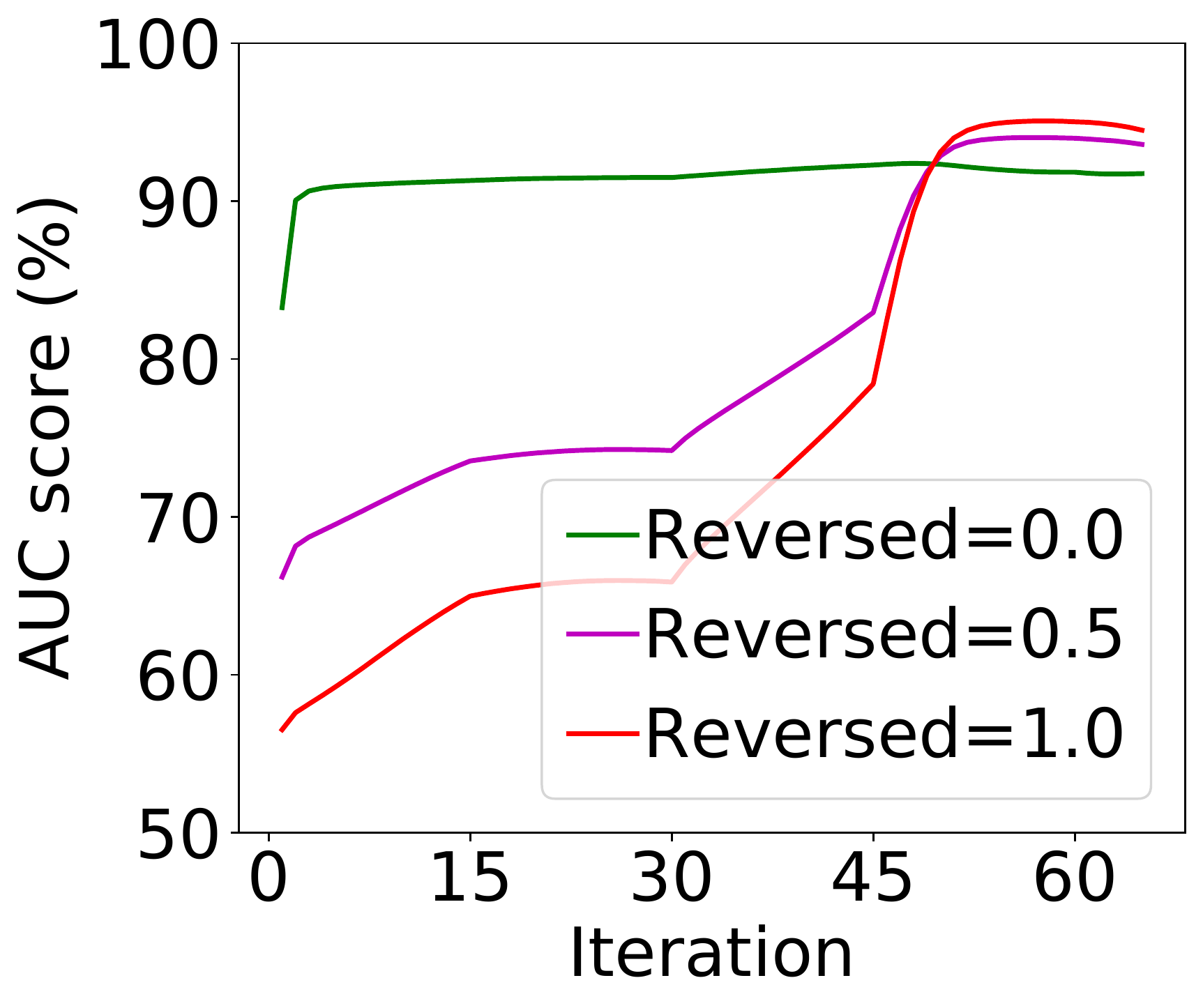} \label{figure5-b}}
\caption{Performance w.r.t. network sparsity and learning curves of DGGAN on Google link prediction task.}
\label{figure5}
\end{figure}

Figure \ref{figure5-a} shows the results with respect to the training ratio of edges on Google dataset. 
One can see that DGGAN consistently and significantly outperforms HOPE and APP across different training ratios.
Moreover, DGGAN still achieves much better performance when the network is very sparse. 
While HOPE and APP extremely suffer from nodes with very low outdegree or indegree as mentioned before. 
It demonstrates that the novel adversarial learning framework DGGAN, which is designed to jointly learn a node’s source and target vectors, can significantly improve the robustness.

Next, we investigate performance change with respect to the training iterations of the discriminator $D$.
Recall that we set the parameter of discriminator training iterations per epoch $n^D = 15$.
Figure \ref{figure5-b} shows the converging performance of DGGAN on Google dataset with different percentage of reversed positive edges of test set (results on other datasets show similar trends and are not included here).
With the increase of iterations of $D$, the performance of Reversed=0.0 (i.e. random negative edges in test set) keeps stable first, and then slightly increases.
Besides, the training curve trend of Reversed=1.0 (i.e. all positive edges except bi-directional edges are reversed to create negative edges in test set) changes every 15 iterations (i.e. one epoch).
Note that the training curve trend of Reversed=1.0 rises gently during second epoch (i.e. iteration $[16, 30]$) for the generator $G$ still been poorly trained at the moment.
The trend rises steep in the following epoch for $G$ being able to generate close-to-real fake samples.

\section{Conclusion}

In this paper, we proposed DGGAN, a novel directed graph embedding framework based on GAN.
Specifically, we designed two generators which generate fake source neighborhood and target neighborhood for each node directly from same continuous distribution.
With the jointly learning framework, the two generators can be mutually enhanced, which renders the proposed DGGAN generalized for various graphs and more robust to learn node embeddings.
The experimental results on four real-world directed graph datasets demonstrated that DGGAN consistently and significantly outperforms various state-of-the-arts on link prediction, node classification, and graph reconstruction tasks.


\section{ Acknowledgments}
Jianxin Li is the corresponding author.
This work was supported by grants from the Natural Science Foundation of China (61872022, U20B2053 and 62002007), State Key Laboratory of Software Development Environment (SKLSDE-2020ZX-12), Natural Science Foundation of China-Guangdong Province (2017A030313339) and CAAI-Huawei MindSpore Open Fund.

\bibliography{aaai21}

\end{document}